\documentclass[prl,aps,twocolumn]{revtex4}
\usepackage{epsf}

 \newcommand{\np}{${\bf NP}$}
 \newcommand{\p}{${\bf P}$}
 
\begin{document} 

\title{Computational complexity and fundamental limitations to fermionic quantum Monte Carlo simulations} 

\author{Matthias Troyer,$^{1}$ Uwe-Jens Wiese$^{2}$}
\affiliation{$^{1}$Theoretische Physik, ETH Z\"urich, CH-8093 Z\"urich, Switzerland}
\affiliation{$^{2}$Institut f\"ur theoretische Physik, Universit\"at Bern, Sidlerstrasse 5, CH-3012 Bern, Switzerland}

\begin{abstract}
Quantum Monte Carlo simulations, while being efficient for bosons, suffer from the ``negative sign problem'' when applied to fermions -- causing an exponential increase of the computing time with the number of particles. A polynomial time solution to the sign problem is highly desired since it would provide an unbiased and numerically exact method to simulate correlated quantum systems. Here we show, that such a solution is almost certainly unattainable by proving that the sign problem is  \np-hard, implying that a generic solution of the sign problem would also solve all problems in the complexity class \np\, (nondeterministic polynomial) in polynomial time.
\end{abstract}
\maketitle 

Half a century after the seminal paper of Metropolis {\it et al.}~\cite{Metropolis} the Monte Carlo method has widely been established as one of the most important numerical methods and as a key to the simulation of many-body problems. Its main advantage is that it allows phase space integrals for many-particle problems, such as thermal averages, to be evaluated in a time that scales only polynomially with the particle number $N$ although the configuration space grows exponentially with $N$. This enables the accurate simulation of large systems with millions of particles. 

Monte Carlo simulations of quantum systems, such as fermions, bosons, or quantum spins, can be performed after mapping the quantum system to an equivalent classical system. For fermionic or frustrated models this mapping may yield configurations with negative Boltzmann weights, resulting in an exponential growth of the statistical error and hence the simulation time with the number of particles, defeating the advantage of the Monte Carlo method. A polynomial time solution of this ``sign problem'' of negative weights would revolutionize electronic structure calculations by providing an unbiased and approximation-free method to study correlated fermionic systems. This would be of invaluable help, for example, in finding the mechanism for high-temperature superconductivity or in determining the properties of dense nuclear matter and quark matter. 

The difficulties in finding polynomial time solutions to the sign problem are reminiscent of the apparent impossibility to find polynomial time algorithms for nondeterministic polynomial (\np)-complete decision problems, which could be solved in polynomial time on a hypothetical non-deterministic machine, but for which no polynomial time algorithm is known for deterministic classical computers. A hypothetical non-deterministic machine can always follow both branches of an if-statement simultaneously, but can never merge the branches again. It can, equivalently, be viewed as having exponentially many processors, but without any communication between them. In addition, it must be possible to check a positive answer to a problem in \np~on a classical computer in polynomial time.

Many important computational problems in the complexity class \np, including the traveling salesman problem and the problem of finding ground states of spin glasses have the additional property of being \np-hard, forming the subset of \np-complete problems, the hardest problems in \np. A problem is called \np-hard if any problem in \np~can be mapped onto it  with polynomial complexity. Solving an \np-hard problem is thus equivalent to solving any problem in \np, and finding a polynomial time solution to any of them would have important consequences for all of computing as well as the security of classical encryption schemes. In that case all problems in \np~could be solved in polynomial time, and hence \np=\p.

As no polynomial solution to any of the \np-complete problems was found despite decades of intensive research, it is generally believed that \np$\ne$\p~and no deterministic polynomial time algorithm exists for these problems.  The proof of this conjecture remains as one of the unsolved millennium problems of mathematics for which the Clay Mathematics Institute has offered a prize of one million US\$ \cite{clay}.
In this Letter we will show that the sign problem is \np-hard, implying that unless the  \np$\ne$\p~ conjecture is disproven there exists no generic solution of the sign problem.

Before presenting the details of our proof, we will give a short introduction to classical and quantum Monte Carlo simulations and the origin of the sign problem. In the calculation of the phase space average of a quantity $A$, instead of directly evaluating the sum
 
\begin{equation}
\langle A \rangle = \frac{1}{Z}\sum_{c\in\Omega} A(c)p(c) \;,\quad Z=\sum_{c\in\Omega} p(c) ,
\label{eq:int}
\end{equation}
over a high-dimensional space $\Omega$ of configurations $c$, a classical Monte Carlo method chooses a set of $M$ configurations $\{c_i\}$ from $\Omega$, according to the distribution $p(c_i)$. The average is then approximated by the sample mean
\begin{equation}
\langle A \rangle \approx \overline{A}=\frac{1}{M}\sum_{i=1}^M A(c_i),
\end{equation}
within a statistical error $\Delta A = \sqrt{{\rm Var} A (2\tau_A+1)/M} $, where ${\rm Var}A$ is the variance of $A$ and the integrated autocorrelation time $\tau_A$ is a measure of the autocorrelations of the sequence $\{A(c_i)\}$. In typical statistical physics applications, $p(c)=\exp(-\beta E(c))$ is the Boltzmann weight, $\beta=1/k_B T$ is the inverse temperature, and $E(c)$ is the energy of the configuration $c$. 

Since the dimension of configuration space $\Omega$ grows linearly with the number $N$ of particles, the computational effort for the direct integration Eq.\ (\ref{eq:int}) scales exponentially with the particle number $N$. Using the Monte Carlo approach the same average can be estimated to any desired accuracy in {\it polynomial time}, as long as the autocorrelation time $\tau_A$ does not increase faster than polynomially with $N$.

In a quantum system with Hamilton operator $H$, instead of an integral like Eq.\ (\ref{eq:int}), an operator expression 
\begin{equation}
\langle A\rangle = \frac{1}{Z}{\rm Tr}[A\exp(-\beta H)]\;,\quad Z={\rm Tr}\exp(-\beta H)
\end{equation}
needs to be evaluated in order to calculate the thermal average of the observable $A$ (represented by a self-adjoint operator).  
Monte Carlo techniques can again be applied to reduce the exponential scaling of the problem, but only after mapping the quantum model to a classical one. One approach to this mapping\cite{independence} is a Taylor expansion \cite{sse}:
\begin{eqnarray}
Z &=& {\rm Tr}\exp(-\beta H) = \sum_{n=0}^{\infty} \frac{(-\beta)^n}{n!}{\rm Tr}H^n  \\
&=& \sum_{n=0}^{\infty}\sum_{i_1,...,i_n}\frac{(-\beta)^n}{n!}\langle i_1|H|i_2\rangle\langle i_2|H|i_3\rangle\cdots\langle i_n|H|i_1\rangle \nonumber\\
&\equiv& \sum_{n=0}^{\infty} \sum_{i_1,...,i_n} p(i_1,...,i_n) \equiv \sum_c p(c) ,\nonumber
\end{eqnarray}
where for each order $n$ in the expansion we insert $n$ sums over complete sets of basis states $\{|i\rangle\}$.  The ``configurations'' are sequences $c=(i_1,...,i_n)$ of $n$ basis states and we  define the weight $p(c)$ by the corresponding product of matrix elements of $H$ and the term $(-\beta)^n/n!$.
With a similar expansion for ${\rm Tr}[A\exp(-\beta H)]$ we obtain an expression reminiscent of classical problems:
\begin{equation}
\langle A\rangle = \frac{1}{Z}{\rm Tr}[A\exp(-\beta H)]= \frac{1}{Z}\sum_c A(c)p(c).
\label{eq:quant}
\end{equation}

If all the weights $p(c)$ are positive, standard Monte Carlo methods can be applied, as it is the case for non-frustrated quantum magnets and bosonic systems. In fermionic systems \cite{frustrated} negative weights $p(c)<0$ arise from the Pauli exclusion principle, when along the sequence $|i_1\rangle\rightarrow |i_2\rangle\rightarrow\cdots\rightarrow |i_n\rangle\rightarrow |i_1\rangle$  two fermions are exchanged, as shown in Fig.\ 1.

\begin{figure}
\begin{center}
\epsfxsize=4cm\epsffile{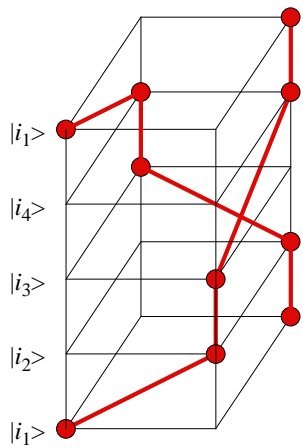}
\end{center}
\caption{A configuration of a fermionic lattice model on a 4-site square. The configuration has negative weight, since two fermions are exchanged in the sequence $|i_1\rangle\rightarrow|i_2\rangle\rightarrow|i_3\rangle\rightarrow|i_4\rangle\rightarrow|i_1\rangle$. World lines connecting particles on neighboring slices are drawn as thick lines.}
\end{figure}

The standard way of dealing  with the negative weights of the fermionic system is to sample with respect to the bosonic system by using the absolute values of the weights $|p(c)|$ and to assign the sign $s(c)\equiv {\rm sign}\, p(c)$ to the quantity being sampled:
\begin{eqnarray}
\langle A\rangle &=& \frac{\sum_c A(c)p(c)}{\sum_c p(c)} \\ &=& \frac{\sum_c A(c)s(c)|p(c)|\left/\sum_c |p(c)|\right.}{\sum_c s(c)|p(c)|\left/\sum_c |p(c)|\right.} \equiv \frac{\langle A s\rangle'}{\langle s \rangle'} .\nonumber
\end{eqnarray}

While this allows Monte Carlo simulations to be performed, the errors increase exponentially with the particle number $N$ and the inverse temperature $\beta$. To see this, consider the mean value of the sign $\langle s \rangle = Z/Z'$, which is just the ratio of the partition functions of the fermionic system $Z=\sum_cp(c)$ with weights $p(c)$ and the bosonic system used for sampling with $Z'=\sum_c |p(c)|$. As the partition functions are exponentials of the corresponding free energies, this ratio is an exponential of the differences $\Delta f$ in the free energy densities:$\langle s \rangle = Z/Z'=\exp(-\beta N \Delta f)$. As a consequence, the relative error $\Delta s / \langle s \rangle$ increases exponentially with increasing particle number and inverse temperature:
\begin{equation}
\frac{\Delta s}{\langle s \rangle} 
=\frac{\sqrt{\left(\langle s^2 \rangle-\langle s \rangle^2\right)/M}}{\langle s \rangle} 
=\frac{\sqrt{1-\langle s \rangle^2}}{\sqrt{M}\langle s \rangle} 
\sim 
\frac{e^{\beta N \Delta f}}{\sqrt{M}} .
\end{equation}

Similarly the error for the numerator in Eq.\ (7) increases exponentially and the time needed  to achieve a given relative error scales exponentially in $N$ and $\beta$.

In order to avoid any misconception about what would constitute a ``solution'' of the sign problem, we start by giving a precise 
definition:

\begin{itemize}
\item 
A quantum Monte Carlo simulation to calculate a thermal average $\langle A \rangle$ of an observable $A$ in a quantum system with Hamilton operator $H$ is defined to suffer from a {\it sign problem} if there occur negative weights $p(c)<0$ in the classical representation as given by Eq.\ (\ref{eq:quant}).
\item 
The related {\it bosonic system} of a fermionic quantum system is defined as the system where the weights $p(c)$ are replaced by their absolute values $|p(c)|$, thus ignoring the minus sign coming from fermion exchanges:
\begin{equation}
\langle A\rangle'  = \frac{1}{Z'}\sum_c A(c)|p(c)| .
\label{eq:bos}
 \end{equation}
\item 
An algorithm for the stochastic evaluation of  a thermal average such as Eqns. (\ref{eq:quant}) or (\ref{eq:bos}) is defined to be of {\it polynomial complexity} if the computational time $t(\epsilon,N,\beta)$ needed to achieve a relative statistical error $\epsilon=\Delta A/\langle A\rangle$ in the evaluation of the average $\langle A \rangle$ scales polynomially with the system size $N$ and inverse temperature $\beta$, i.e.\ if there exist integers $n$ and $m$ and a constant $\kappa<\infty$ such that
\begin{equation}
t(\epsilon,N,\beta) < \kappa\epsilon^{-2} N^n \beta^m .
\end{equation}
\item 
For a quantum system that suffers from a sign problem for an observable $A$, and for which there exists a polynomial complexity algorithm for the related bosonic system Eq.\ (\ref{eq:bos}), we define a {\it solution of the sign problem} as an algorithm of polynomial complexity to evaluate the thermal average $\langle A\rangle$.
\end{itemize}

It is important to note that we only worry about the sign problem if the bosonic problem is easy (of polynomial complexity) but the fermionic problem hard (of exponential complexity) due to the sign problem. If the bosonic  problem is already hard, e.g. for spin glasses \cite{spinglass}, the sign problem will not increase the complexity of the problem.
Also, changing the representation so that the sum in Eq.\ (\ref{eq:quant}) contains only positive terms $p(c)\ge0$ is not sufficient to solve the sign problem if the scaling remains exponential, since then we just map the sign problem to another exponentially hard problem. Only a polynomial complexity algorithm counts as a solution of the sign problem.

At first sight such a solution seems feasible since the sign problem is not an intrinsic property of the quantum model studied but is representation-dependent: it depends on the choice of basis sets $\{|i\rangle\}$, and in some models it can be solved by a simple local basis change  \cite{hatano}. 
Indeed, when using the eigenbasis in which the Hamilton operator $H$ is diagonal, there will be no sign problem. This diagonalization of the Hamilton operator is, however, no solution of the sign problem since its complexity is exponential in the number of particles $N$.

We now construct a quantum mechanical system for which the calculation of a thermal average provides the solution for one and thus all of the \np-complete problems. This system exhibits a sign problem, but the related bosonic problem is easy to solve. Since, for this model, a solution of the sign problem would provide us with a polynomial time algorithm for an \np-complete problem, the sign problem is \np-hard. Of course, it is expected that the corresponding thermal averages cannot be calculated in polynomial time and the sign problem thus cannot be solved. Otherwise we would have found a polynomial time algorithm for the \np-complete problems and would have shown that \np=\p.

The specific  \np-complete problem we consider  \cite{spinglass} is to determine whether a state with energy less than or equal to a bound $E_0$ exists for a classical three-dimensional Ising spin glass with Hamilton function
\begin{equation}
H =- \sum_{\langle j,k \rangle} J_{jk}\sigma_j\sigma_k.
\end{equation}
Here the spins $\sigma_j$ take the values $\pm1$, and the couplings $J_{jk}$ between nearest neighbor lattice points $j$ and $k$ are either $0$ or $\pm J$. 

This problem is in the complexity class \np~since the non-deterministic machine can evaluate the energies of all configurations $c$ in polynomial time and test whether there is one with $E(c)\le E_0$. In addition, the validity of a positive answer (i.e.\ there is a configuration $c$) can be tested on a deterministic machine by evaluating the energy of that configuration. The evaluation of the partition function $Z=\sum_c\exp(-\beta E(c))$ is, however, not in \np~since the non-deterministic machine cannot perform the sum in polynomial time.

This question whether there is a state with energy  $E(c)\le E_0$ can also be answered in a Monte Carlo simulation by calculating the average energy of the spin glass at a large enough inverse temperature $\beta$. Since the energy levels are discrete with spacing $J$ it can easily  be shown that by choosing
an inverse temperature $\beta J \ge N\ln 2 +\ln(12 N)$ the thermal average of the energy will be less than $E_0+J/2$ if at least one configuration with energy $E_0$ or less exists, and larger than $E_0+J$ otherwise \cite {bpp}.

In this classical Monte Carlo simulation, the complex energy landscape, created by the frustration in the spin glass (Fig.\ 2a), exponentially suppresses the tunneling of the Monte Carlo simulation between local minima at low temperatures. The autocorrelation times and hence the time complexity of this Monte Carlo approach are exponentially large $\tau \propto \exp(a N)$, as expected for this \np-complete problem.

\begin{figure}
\begin{center}
\epsfxsize=6cm\epsffile{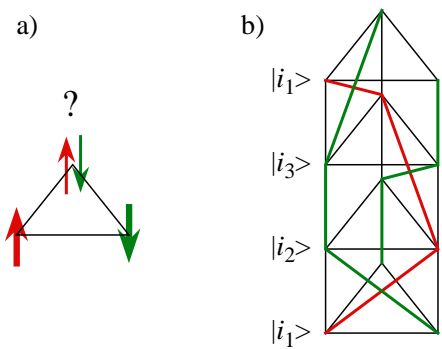}
\end{center}
\caption{a) A classically frustrated spin configuration of three antiferromagnetically coupled spins: no configuration can simultaneously minimize the energy of all three bonds. b) A configuration of a frustrated quantum magnet with negative weights: three antiferromagnetic exchange terms with negative weights are present in the sequence $|i_1\rangle\rightarrow|i_2\rangle\rightarrow|i_3\rangle\rightarrow|i_1\rangle$. Here up-spins  with $z$-component of spin $\sigma_j^z=1$ and down-spins with $\sigma_j^z=-1$ are connected with differently colored world lines.}
\end{figure}

We now map this classical system to a quantum system with a sign problem. We do so by replacing the classical Ising spins by quantum spins. Instead of the common choice in which the classical spin configurations are basis states and the spins are represented by diagonal $\sigma_j^z$ Pauli matrices we choose a representation in which the spins point in the $\pm x$ direction and are represented by $\sigma_j^x$ Pauli matrices:
\begin{equation}
H =- \sum_{\langle j,k \rangle} J_{jk}\sigma^x_j\sigma^x_k,
\end{equation}
Here the random signs of the couplings are mapped to random signs of the off-diagonal matrix elements which cause a sign problem (see Fig.\ 2b). The related bosonic model is the ferromagnet with all couplings $J_{jk}\ge0$ and efficient cluster algorithms with polynomial time complexity are known for this model \cite{evertz}. Since the bosonic version is easy to simulate, the {\it sign problem is the origin of the \np-hardness} of a quantum Monte Carlo simulation of this model. A generic solution of the sign problem would provide a polynomial time solution to this, and thus to all, \np-complete problems, and would hence imply that \np=\p. Since it is generally believed that \np$\ne$\p, we expect that such a solution does not exist.

{\it Conclusions} --
By constructing a concrete model we have shown that the sign problem of quantum Monte Carlo simulations is \np-hard. This does not exclude that a specific sign problem can be solved for a restricted subclass of quantum systems. This was indeed possible using the meron-cluster algorithm \cite{merons} for some particular lattice models. Such a solution must be intimately tied to properties of the physical system and allow an essentially bosonic description of the quantum problem. 
A generic approach like the cancellation idea \cite{cancel} might scale polynomially for some cases but will in general scale exponentially.

In the case of fermions or frustrated quantum magnets, solving the sign problem requires a  mapping to a bosonic or non-frustrated system -- which is, in general, almost certainly impossible for physical reasons. The origin of the sign problem is, in fact, the distinction between bosonic and fermionic systems. The brute-force approach of taking the absolute values of the probabilities means trying to sample a frustrated or fermionic system by simulating a non-frustrated or bosonic one. As for large system sizes $N$ and low temperatures the relevant configurations for the latter are not the relevant ones for the former, the errors are exponentially large.

Given the \np-hardness of the sign problem one promising idea for the simulation of fermionic systems is to use ultra-cold atoms in optical lattices to construct well-controlled and tunable implementations of physical systems, such as the Hubbard model \cite{optical}, and to use these ``quantum simulators'' to study the phase diagrams of correlated quantum systems. But even these  quantum simulators are most likely not a generic solution to the sign problem since there exist quantum systems with exponentially diverging time scales and it is at present not clear whether a quantum computer could solve the \np-complete problems \cite{quantumnp}. 

We like to thank S. Chakravarty, E.\ Farhi, J.\ Goldstone, S.\ Gutmann and G.\ Ortiz for stimulating discussions and acknowledge financial support of the Swiss National Science Foundation. Part of this work was supported by the Kavli Institute for Theoretical Physics and the Aspen Center for Physics.


\begin{thebibliography}{99}
\bibitem{independence} Note that the conclusions drawn in this Letter are independent of the representation.
\end{thebibliography}
\end{document}